\documentclass[11pt, a4paper]{article}
\pdfoutput=1

\usepackage{jcappub}

\usepackage{graphicx}
\usepackage{datetime}
\usepackage{textpos}
\usepackage{booktabs}
\usepackage{subfigure}

\newcommand{\fex}{\textit{e.g.}}

\newcommand{\DNeff}{\Delta N_{\rm eff}}

\title{Cold dark matter plus not-so-clumpy dark relics}

\author[a]{Roberta Diamanti,}
\author[a]{Shin'ichiro Ando,}
\author[b]{Stefano Gariazzo,}
\author[b]{Olga Mena}
\author[a]{and Christoph Weniger}
\affiliation[a]{GRAPPA, Institute of Physics, University of Amsterdam, Science Park 904,
1098 XH Amsterdam, Netherlands}
\affiliation[b]{IFIC, Universidad de Valencia-CSIC, 46071, Valencia, Spain}
\emailAdd{r.diamanti@uva.nl}
\emailAdd{s.ando@uva.nl}
\emailAdd{gariazzo@to.infn.it}
\emailAdd{omena@ific.uv.es}
\emailAdd{c.weniger@uva.nl}

\abstract{Various particle physics models suggest that, besides the (nearly) cold dark matter that accounts for current observations, additional but sub-dominant dark relics might exist.  These could be warm, hot, or even contribute as dark radiation.  We present here a comprehensive study of two-component dark matter scenarios, where the first component is assumed to be cold, and the second is a non-cold thermal relic.  Considering the cases where the non-cold dark matter species could be either a fermion or a boson, we derive consistent upper limits on the non-cold dark relic energy density for a very large range of velocity dispersions, covering the entire range from dark radiation to cold dark matter.  To this end, we employ the latest Planck Cosmic Microwave Background data, the recent BOSS DR11 and other Baryon Acoustic Oscillation measurements, and also constraints on the number of Milky Way satellites, the latter of which provides a measure of the suppression of the matter power spectrum at the smallest scales due to the free-streaming of the non-cold dark matter component.  We present the results on the fraction $f_{\rm ncdm}$ of non-cold dark matter with respect to the total dark matter for different ranges of the non-cold dark matter masses.  We find that the 2$\sigma$ limits for non-cold dark matter particles with masses in the range 1--10 keV are $f_{\rm ncdm}\leq0.29$ (0.23) for fermions (bosons), and for masses in the 10--100 keV range they are $f_{\rm ncdm}\leq0.43$ (0.45), respectively.} 

\begin{document}

\maketitle

\section{Introduction}

Despite lots of efforts during the past decades, the nature of dark matter (DM) remains still unknown (see Ref.~\cite{Bertone:2004pz, Bergstrom:2012fi, Kusenko:2013saa} for extensive reviews).  Usually, it is assumed that the DM is made of heavy, cold thermal relic particles that decoupled from baryonic matter in the very early Universe.  This scenario is commonly dubbed cold dark matter (CDM).  CDM is one of the most abundant ingredients of the standard cosmological model, the $\Lambda$CDM model, where CDM accounts for $\sim$26\% of the current energy density of the Universe.  The remaining part is dominated by dark energy, responsible for the present accelerated expansion of the Universe.  According to the latest measurements of the Cosmic Microwave Background (CMB) from the Planck satellite~\cite{Ade:2015xua}, dark energy accounts for $\sim$69\% of the present total energy density in the Universe.  The $\Lambda$CDM model is extremely successful, being consistent with the majority of current cosmological measurements such as, for instance, the acoustic peaks of CMB and the large scale structure observations~\cite{Ade:2015xua, Anderson:2013zyy}.  Despite its great observational success, there are some pending issues that still need to be understood within the standard $\Lambda$CDM framework.  These are related to various cosmological observations on Galactic and sub-Galactic scales that are not in full agreement with the predictions of the $\Lambda$CDM model (see, \fex,~\cite{Perivolaropoulos:2008ud, Shanks:2004af}).  

Important challenges for the $\Lambda$CDM model are the so-called \emph{too big to fail problem}~\cite{BoylanKolchin:2011dk} and the \emph{Milky Way satellite problem}~\cite{Klypin:1999uc, Moore:1999nt}.  The former refers to the fact that the measured Galactic velocities indicate that dwarf galaxies are hosted by haloes that are less massive than those predicted by numerical simulations based on the $\Lambda$CDM model.  The latter is related to predictions within the $\Lambda$CDM cosmology for the number of DM sub-halos, which is much larger than the observed number of satellite galaxies that orbit close to the Milky Way.  A number of solutions to these two problems have been proposed in the literature (see, \fex, the recent works~\cite{Wang:2016rio, Lovell:2016nkp} and references therein).  Possible avenues range from lowering the total mass of the satellites by different means (\fex, via baryon or supernovae feedback effects~\cite{Sawala:2012cn,Sawala:2015cdf}; see also Ref.~\cite{Fattahi:2016nld}) to modifications of the underlying DM model.\footnote{For another possibility which deals with the cosmological parameter values used for the numerical N-body simulations, see~\cite{Polisensky:2013ppa}.} Focusing on this last solution, the possible modifications to the standard CDM paradigm include self-interacting DM models~\cite{Vogelsberger:2012ku}, interacting DM-radiation models~\cite{Schewtschenko:2015rno} and warm DM particles (such as a sterile neutrino)~\cite{Lovell:2011rd,Lovell:2013ola,Lovell:2015psz}, which attracted recently again more attention due to possible hints in X-ray data~\cite{Bulbul:2014sua, Boyarsky:2014jta}.

\medskip

Here, we consider a modified version of the most economical pure CDM model, allowing for a mixed DM cosmology with an additional dark and inert relic.  An important subset of these models, where today's DM consists of an admixture of cold and warm DM particles, have been dubbed mixed DM (MDM) models; see Refs.~\cite{Maccio':2012uh, Anderhalden:2012qt,Anderhalden:2012jc}.  They are a plausible solution to alleviate the small-scale crisis of the $\Lambda$CDM cosmology, while leaving the predictions from the CDM model at large scales unchanged.  The reason is simple: the particle associated to the second, warm DM component will have a significant free-streaming length, affecting the matter power spectrum on the smallest scales, improving therefore the compatibility with the observations of the local Universe~\cite{Weinberg:2013aya}.

We will study the phenomenology of a broad range of MDM models, where the dominant species is cold, and the sub-dominant species can be warm (see Refs.~\cite{Viel:2005qj,Boyarsky:2008xj,Boyarsky:2008mt} for previous works), hot, or even contribute as a relativistic component (see \fex~Ref.~\cite{Archidiacono:2013fha} and references therein).  More precisely, our goal is to study how cosmological measurements can be used to derive limits on the fraction $f_{\rm ncdm}$ of the non-cold DM (NCDM) component with respect to the total DM, as a function of its mass $m_{\rm ncdm}$.  We will assume that the temperature of the NCDM component is the same as the one of the standard neutrinos, but this does not restrict the scope of our findings (see Sec.~\ref{SEC2}).  We shall exploit the predictions within MDM scenarios of the matter power spectrum $P(k)$ to compute the number of the dwarf satellite galaxies, following~\cite{Schneider:2014rda}.  We shall combine in our analyses the predictions of the number of satellite galaxies with the most recent CMB temperature, polarization and lensing measurements from the Planck satellite and the most recent Baryon Acoustic Oscillation (BAO) data.  Our results will be presented for two typical classes of candidates for the NCDM component: fermions and bosons that froze-out when still relativistic.  Even if we will not consider any specific particle physics model, we will show that our treatment is able to provide constraints on several fermionic or bosonic DM candidates.

One fermionic example could be a sterile neutrino; see Refs.~\cite{Lovell:2016nkp,Lovell:2015psz} for more recent work and Refs.~\cite{Kusenko:2009up, Boyarsky:2012rt,Abazajian:2012ys} for reviews on the subject.  Sterile neutrinos are usually produced as relativistic particles, and therefore their free-streaming can wash out the growth of perturbations at scales of dwarf galaxies and below.  In the very first and simplest model, sterile neutrinos can be produced in the early Universe via mixing with the active neutrinos~\cite{Dodelson:1993je, Dolgov:2000ew, Asaka:2006nq, Seljak:2006qw, Vincent:2014rja}.  However, the sterile neutrino production could be enhanced in the presence of a non-zero lepton asymmetry in the early Universe, or via resonant-production~\cite{Shi:1998km, Abazajian:2001nj, Abazajian:2001vt, Abazajian:2014gza, Bozek:2015bdo, Horiuchi:2015qri}.  Furthermore, sterile neutrinos could also be produced by interactions beyond the SM, for instance by interactions with the Higgs boson, in theories with higher dimensional operators, or in Starobinsky-like inflationary scenarios~\cite{Kusenko:2006rh, Shaposhnikov:2006xi, Anisimov:2008qs, Petraki:2007gq, Shoemaker:2010fg, Bezrukov:2011sz, Bezrukov:2008ut, Gorbunov:2010bn}.  We must underline that most of the models we mentioned do not predict a thermal momentum distribution function for the sterile neutrino, as instead we are assuming in our analysis.

A typical bosonic NCDM candidate is the thermal axion.  Axions are hypothetical elementary particles introduced to resolve the strong CP problem in Quantum Chromodynamics (QCD)~\cite{Peccei:1977hh, Peccei:1977ur, Weinberg:1977ma, Wilczek:1977pj}; see Ref.~\cite{Marsh:2015xka} for review.  Axions can be produced non-thermally by the re-alignment mechanism and, depending on the cosmological scenario, by decaying axion strings and domain walls, contributing to the CDM of the Universe~\cite{Preskill:1982cy, Abbott:1982af, Dine:1982ah, Turner:1990uz, Sikivie:2006ni, Wantz:2009it, Kawasaki:2013ae, Hiramatsu:2012gg, Visinelli:2009zm, Visinelli:2014twa, DiValentino:2014zna}, but we shall not focus here on this case.  On the contrary, light, sub-eV axions can be copiously produced in the early Universe via thermal processes, behaving as extra-hot DM components, together with the three standard relic neutrinos; see Refs.~\cite{Archidiacono:2013cha, DiValentino:2016ikp, Archidiacono:2015mda,DiValentino:2015zta, Giusarma:2014zza, DiValentino:2013qma} for current and forecasted future cosmological constraints on these models.

These are just a few examples for the NCDM component that we will cover in our analysis.  Our results, indeed, will be derived in a much more extended parameter space than the sterile neutrino and thermal axion ones.

\medskip

The structure of this paper is as follows: in Sec.~\ref{SEC2}, we briefly describe relevant aspects of the standard cosmological model, and we introduce the second DM component.  Section~\ref{DATA} contains a detailed description of the adopted datasets and the tools that are used in our numerical analyses, including the computation of the number of the dwarf satellite galaxies.  We present our results in Sec.~\ref{RES}, and we draw our conclusions in Sec.~\ref{CON}.

\section{The standard cosmological scenario and a second dark matter component}
 \label{SEC2}
 
The $\Lambda$CDM model has six parameters: the energy density of CDM, $\Omega_{cdm}$; the baryon energy density, $\Omega_b$; the reionization optical depth, $\tau$; the angular scale of acoustic peaks, $\theta$; the amplitude and the tilt of the power spectrum of the primordial curvature perturbations, $A_s$ and $n_s$.

The amount of energy density of relativistic species in the early Universe is usually defined as the sum of the photon contribution, $\rho_\gamma$, plus the contribution of all the other relativistic species, parametrized through the effective number of relativistic degrees of freedom, $N_{\rm eff}$:
\begin{equation}\label{RHO}
  \rho_{rad} = \left[ 1 + \frac{7}{8} \left(\frac{4}{11}\right)^{4/3} \, N_{\rm eff} \right] \rho_\gamma = [1 + 0.227 N_{\rm eff} ]\rho_\gamma \,~.
\end{equation}
$N_{\rm eff}$ is defined as the ratio of the energy density of all relativistic species, $\rho_x$ (which includes standard neutrinos plus any other NCDM component in its relativistic regime), to that of photons:
\begin{equation}\label{NEFF}
  N_{\rm eff} = \left(\frac{8}{7}\right)\left(\frac{11}{4}\right)^{4/3}\, \frac{\rho_{x}}{\rho_\gamma} \,.
\end{equation}

In the standard scenario, the canonical value $N_{\rm eff} = 3.046$ corresponds to the three active neutrino contribution, after considering effects related to non-instantaneous neutrino decoupling~\cite{Mangano:2005cc,deSalas:2016ztq}.  Deviations of $N_{\rm eff}$ from its standard value may indicate that the thermal history of the active neutrinos is different from what we expect from the Standard Model (SM) of particle physics (\fex, additional relativistic particles that may be present in the Universe).  The extra dark radiation component is parametrized via $\DNeff \equiv N_{\rm eff} - 3.046$.

The most recent measurements from the Planck satellite using both temperature and polarization, combined also with BAO data, give $N^{\rm CMB}_{\rm eff} = 3.04 \pm 0.18$~\cite{Ade:2015xua} (see also Refs.~\cite{Feeney:2013wp,Verde:2013cqa,Rossi:2014nea,Giusarma:2014zza,DiValentino:2015sam,DiValentino:2016ikp,Archidiacono:2016kkh} for other recent constraints, and the review~\cite{Archidiacono:2013fha} that carefully analyses the impact of $N_{\rm eff}$ on the different cosmological observables).

\medskip

The phase space distribution of particles in thermal equilibrium that decoupled when still relativistic reads
\begin{equation}\label{F}
  f(\vec{p}) = \frac{1}{e^{(E - \mu)/T} \pm 1} \,,
\end{equation}
where the sign $+$($-$) is for fermions (bosons), $T$ is their temperature and $\mu$ the chemical potential.  We will fix the temperature of the NCDM component today to $T_{\rm ncdm} = T_\nu = 0.716 T_{\rm cmb}$, the temperature of the active neutrinos, and study the NCDM as function of the NCDM mass $m_{\rm ncdm}$ and the fractional contribution to DM (see below).  Even if we consider a specific value of $T_{\rm ncdm}$, we are not restricting ourselves to a fixed scenario, since what is relevant for the cosmological calculations is the ratio $m_{\rm ncdm}/T_{\rm ncdm}$.  This means that the results that we will find for a particle with mass $m_{\rm ncdm}$ and temperature $T_{\rm ncdm} = T_\nu$ can be easily translated into constraints for a model where the NCDM particle is described by any other temperature $T'$ by rescaling the mass accordingly, i.e. $m'=m_{\rm ncdm}T'/T_{\rm ncdm}$.

We assume that the NCDM component freezes out while still being relativistic ($E=p$), and with zero chemical potential ($\mu=0$).  In that case, the functional form of the momentum distribution is conserved at all later times, with a temperature re-shifting as $T\propto 1+z$ as the Universe expands~\cite{Kolb:1990vq}.  The number density, or equivalently the normalization of the distribution function in Eq.~\eqref{F}, is fixed by requiring that the energy density of the NCDM component equals $\rho_\text{ncdm}$.  

\medskip

We \textit{define} the fractional amount of NCDM as
\begin{equation}\label{1}
  f_{\rm ncdm} \equiv \frac{ \Omega_{\rm ncdm}} {\Omega_{\rm cdm} + \Omega_{\rm ncdm}}~,
\end{equation}
where $\Omega_x \equiv \rho_x/\rho_c$, $\rho_{\rm cdm}$ and $\rho_{\rm ncdm}$ are the mass-energy densities of the CDM and NCDM components, respectively, and $\rho_c$ is the critical density of the Universe.  Note that the total energy density of DM today is, strictly speaking, \textit{not} exactly given by $\Omega_{\rm cdm} + \Omega_{\rm ncdm}$, since $\Omega_{\rm ncdm}$ can be potentially relativistic in our scenario.  However, we note that for all cases of interest, $\Omega_{\rm dm} \approx \Omega_{\rm cdm} + \Omega_{\rm ncdm}$ will be approximately correct, since the fraction of dark radiation that can contribute to the energy density today is strongly constrained (as we will explicitly see below).

\section{Observational constraints}
\label{DATA}

\subsection{Cosmological measurements}

We consider the CMB measurements of the most recent Planck data release~\cite{Ade:2015xua}, using the full temperature power spectrum at multipoles $2 \leq l \leq 2500$ (\texttt{Planck TT}) and the polarization power spectra in the range $2 \leq l \leq 29$ (\texttt{lowP}).  We also include the information on the gravitational lensing power spectrum estimated from the CMB trispectrum analysis, as implemented in the Planck lensing likelihood described in Ref.~\cite{Ade:2015zua}.  Here, we follow a very conservative approach and neglect the small-scale polarization measurements (i.e., the so-called \texttt{highP}), as there could be still some level of systematics contamination~\cite{Ade:2015xua}.  In order to perform our numerical analyses, we have made use of the publicly available Planck likelihoods~\cite{Aghanim:2015xee}.\footnote{The likelihood codes are publicly available at the Planck Legacy Archive (\url{http://pla.esac.esa.int/pla/}).} We refer to the combination of the data above described as the ``CMB dataset''.

We also consider the BAO measurements from several experiments: from 6dFGS~\cite{Beutler:2011hx} at redshift $z = 0.1$, from the SDSS Main Galaxy Sample (MGS)~\cite{Ross:2014qpa} at redshift $z_{\rm eff} = 0.15$, and from the BOSS experiment Data Release 11 (DR11) using both the results from the LOWZ and CMASS samples~\cite{Anderson:2013zyy} at redshift $z_{\rm eff} = 0.32$ and $z_{\rm eff} = 0.57$, respectively.

\subsection{Dwarf spheroidal number counts}

Besides including the above CMB and BAO measurements, we derive constraints from the number of dwarf satellite galaxies in the Milky Way.  Their distribution and number is a probe of the matter perturbations at sub-Mpc scales, which can potentially be affected by the free streaming of the NCDM component in our model.

We estimate the number of satellites predicted for a given MDM scenario following the procedure described in Refs.~\cite{Schneider:2014rda, Schneider:2016uqi}, where the authors use a relation based on the conditional mass function normalized to the N-body simulation results. The use of this formalism in the MDM scenarios explored here is justified by the recent work performed in Ref.~\cite{Murgia:2017lwo}, where it has been explicitely shown that there is a very good agreement between the theoretical description of the mass function and the devoted N-body simulations within MDM models carried out in Ref.~\cite{Murgia:2017lwo}.

The expected number of the dwarf satellite galaxies can be calculated integrating the following quantity~\cite{Schneider:2014rda, Schneider:2016uqi}:
\begin{equation}
  \label{eq:sat}
  \frac{dN_{sat}}{d \ln M_{sat}} = \frac{1}{C_n} \, \frac{1}{6 \pi^2} \, \left(\frac{M_{hh}}{M_{sat}}\right) \, \frac{ P(1/R_{sat}) }{R^3_{sat} \sqrt{2\pi(S_{sat} - S_{hh})}}\,~,
\end{equation}
where $P(1/R)$ is the matter power spectrum, $hh$ stands for ``\emph{host halo}'' (the Milky Way in our case), $sat$ stands for ``\emph{satellites},'' and $C_n = 45$ is a number which mimics the results of N-body simulations.  In the rest of the paper we assume $M_{hh}=1.77 \cdot 10^{12} h^{-1}M_\odot$ for the Milky Way (MW) and we take $N_{sat,0}=10^{8} h^{-1}M_\odot$ as a lower integration limit for the calculation of the number of satellite galaxies.  Hence, we only consider satellite galaxies above that mass~\cite{Schneider:2014rda}.

The parameters $R_i$, $S_i$ and $M_i$ ($i=sat, hh$) are, respectively, the radius, the variance and the mass of the satellite galaxies or of the host-halo, defined as
\begin{equation}\label{eq:s_m_halo}
  S_i(M) = \frac{1}{2\pi^2} \int_0^\infty k^2 P(k) W^2(k|M) dk \,,  \quad M_i = \frac{4\pi}{3} \, \Omega_m \rho_c (c R_i)^3 \,~.
\end{equation}
The number $c = 2.5$ is fixed in order to give the best match to N-body simulations.  This approach is based on a re-derivation of the Press \& Schechter \cite{Press:1973iz} mass function.  For the calculation of the mass and the variance, we use a \emph{k-sharp filter} approach.  This filter cuts all the scales $k$ below the cut-off scale $1/R_{sat}$ and it is written in terms of the \emph{window function} $W(k|M)$ that enters the above equation~\eqref{eq:s_m_halo}.  The window function reads as follows:
\begin{equation}
  W(k|M) =
  \begin{cases}
    1, \quad \text{if} \quad k \leq k_s(M)  \,;\\
    0, \quad \text{if} \quad k > k_s(M) \,,\\
  \end{cases}
\end{equation}
where $k_s(M)$ is the cut-off scale as a function of the mass.\footnote{In the standard extended Press \& Schechter~\cite{Press:1973iz} formalism, the window function is chosen to be a top hat in real space.  In Fourier space, it is $W(k|M) = [3 \sin(kR) - kR\cos(kR)]/(kR)^3$, where the relation between the mass $M$ and the filtering scale $R$ reads $M = \frac{4}{3} \pi \rho_c R^3$.  This implies that the mass $M$ and the filtering scale $R$ are related.  This relation is also maintained with the choice of a sharp-k filter~\cite{Lovell:2015psz}.}

The computation of the number of the dwarf spheroidal galaxies enters in a likelihood function labelled `SAT'.  As for the CMB dataset, for which we do not consider Planck polarization data at high multipoles, we also follow here a very conservative approach.  We define the satellite likelihood as a half-Gaussian with mean $N_{\rm{sat}}^{}=61$ and standard deviation $\sigma_{N_{\rm{sat}}}^{}=13$.\footnote{The choice of $N_{\rm{sat}}^{}=61$ is motivated by Refs.~\cite{Schneider:2016uqi, Polisensky:2010rw}, in which the authors add to the eleven standard satellites the fifteen observed by SDSS, after having corrected for the limited sky coverage of the SDSS catalogue ($f_{\rm sky}\simeq 0.28$), resulting in $\sim 61\pm 13$ satellite galaxies.  The error only accounts for the SDSS sample, for which we assume Poisson statistics, see Ref.~\cite{Polisensky:2010rw}.} In other words, we only consider satellite galaxy bounds when the predicted number of satellite galaxies within a given model is \textit{below} the mean number of galaxy satellites that are (expected to be) observed, with the present observations ($N_{\rm{sat}}^{}=61$) representing only a lower limit.  This assumes the plausible scenario that not all dwarf spheroidal galaxies in the relevant mass range have been found up to now, and that the number of satellite galaxies might increase by ongoing and/or future searches.

\medskip

\begin{figure}[ht]
  \begin{center}
    \includegraphics[width=0.6\linewidth]{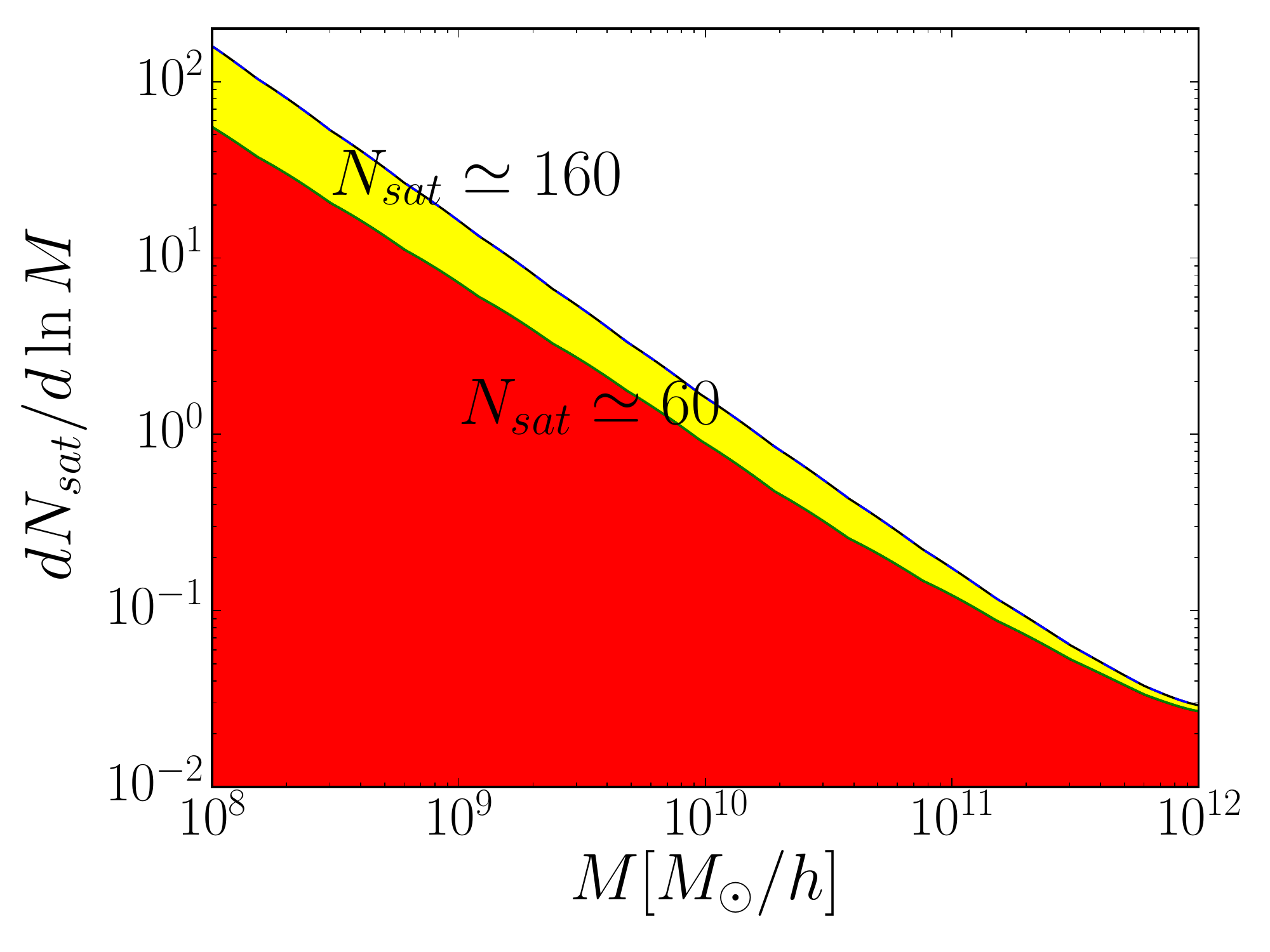}
  \end{center}
  \caption{Estimate of the number of satellites with mass $M$ inside a host halo with mass $M_{hh}= 1.77 \cdot 10^{12} h^{-1}M_\odot$, for two different MDM models.  The thin green solid line which separates the two coloured regions refers to a NCDM component with fraction $f_{\rm ncdm} = 0.25$ and mass $m_{\rm ncdm} = 10^3$ eV, corresponding to a total of $\simeq60$ dwarf spheroidal galaxies satellites.  The black line which limits the upper coloured region refers to a model where the fraction is $f_{\rm ncdm} = 0.9$ and the mass is $m_{\rm ncdm} = 10^5$ eV, corresponding to a total of $\simeq$ 160 dwarf spheroidal galaxies.}
  \label{fig:dn_dlnM}
\end{figure}

In Fig.~\ref{fig:dn_dlnM}, we show the derivative $dN_{sat}/d \ln M$ of the number of satellite galaxies, see Eq.~\eqref{eq:sat}, versus the satellite mass $M$, for a Milky Way-like host halo with mass $M_{hh} = 1.77 \cdot 10^{12} h^{-1}M_\odot$.  We show two cases.  The first case depicts a situation where the mass of the NCDM particle is large ($m_{\rm ncdm} = 10^5$ eV) and it constitutes almost all the DM ($f_{\rm ncdm} = 0.9$), resembling therefore the standard $\Lambda$CDM picture.  The other case is for smaller values of the mass and the fraction of the NCDM species, having $m_{\rm ncdm} = 10^3$ eV and $f_{\rm ncdm} = 0.25$, respectively.  For the $\Lambda$CDM-like scenario we get $N_{\textrm{sat}}\sim 160$, while for the second scenario we obtain $N_{\textrm{sat}}\sim 60$, providing naively a better agreement with observations.  However, we are conservatively imposing that our model must reproduce \emph{at least} the number of observed dwarf satellite galaxies, penalizing only those cases for which the number of satellite galaxies is \emph{smaller} than the observed one.  Hence, both of the exemplary MDM scenarios will be equally allowed by the `SAT' likelihood, since future measurements may detect more of these objects.

\subsection{Boltzmann code and scanner}

We use \texttt{CLASS}, a Boltzmann solver code that calculates the evolution of matter perturbations in the Universe and evaluates the CMB and BAO observables~\cite{Lesgourgues:2011re}.  The tool used for the computation of the likelihoods is \texttt{Montepython}~\cite{Audren:2012wb}, that we use in junction with \texttt{Multinest}, an efficient and robust Bayesian inference tool for cosmology and particle physics~\cite{Feroz:2008xx}.  The varying cosmological parameters are the six basic $\Lambda$CDM parameters introduced in Sec.~\ref{SEC2}, plus the mass of the NCDM component, $m_{\rm ncdm}$ and its fraction, $f_{\rm ncdm}$.  For the standard $\Lambda$CDM cosmological parameters we use flat priors, while for the mass and the fraction of the NCDM component we use flat priors on their logarithms, having $\log_{10} (m_{\rm ncdm}/\text{eV}) \in [-5 ; 5]$ and $\log_{10} f_{ncdm} \in [-6 ; 0]$.  Notice that this is a very wide region for the NCDM properties and therefore there are parts in which our results will overlap with other complementary analyses.  That would be the case for the bosonic axion, where astrophysical constraints apply to a certain region of the parameter space~\cite{Raffelt:2006cw}, and also for the fermionic sterile neutrino, for which oscillation searches discard values smaller than about  1~eV~\cite{Gariazzo:2015rra}.  However, we are focusing here on the constraining power of cosmological tools alone for a wide range of models.  We do not aim to study specific particle physics candidates for which a combination of all the possible available measurements may further restrict the NCDM parameter region.

\medskip

\begin{figure}[t]
  \begin{center}
    \includegraphics[width=0.7\linewidth]{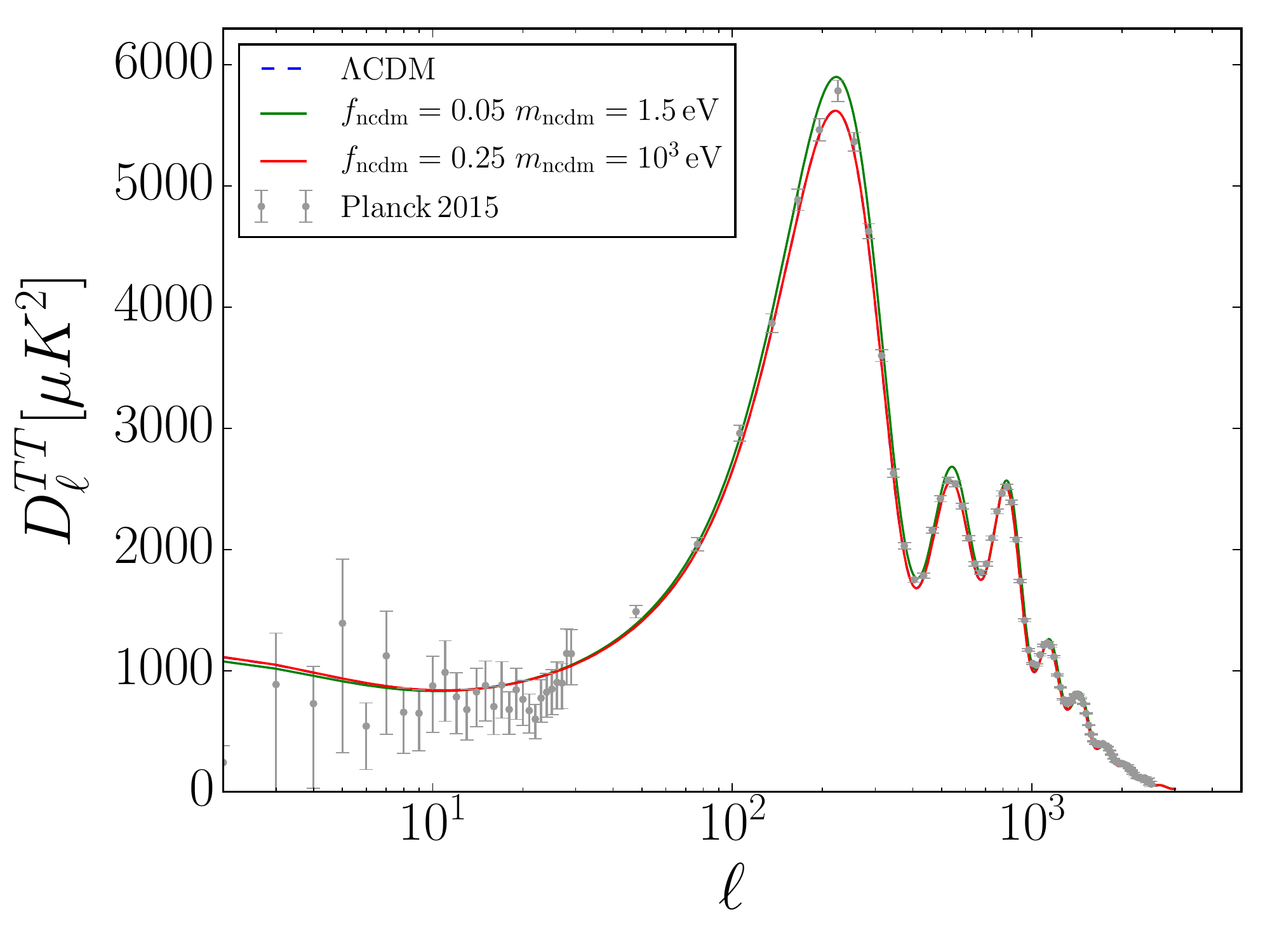}
  \end{center}
  \caption{Angular power spectrum for two different models.  The green (red) solid line refers to a NCDM component with fraction $f_{\rm ncdm} = 0.05$ ($f_{\rm ncdm} = 0.25$) and mass $m_{\rm ncdm} = 1.5$~eV ($m_{\rm ncdm} = 10^3$~eV).  The standard $\Lambda$CDM predictions are depicted by a blue dashed line, that is not visible because it coincides with the red one.  The data points and errors are from the Planck 2015 data release~\cite{Ade:2015xua}. All the cosmological parameters are kept fixed for the three cases illustrated here, except for $f_{\rm ncdm}$ and $m_{\rm ncdm}$. In particular, in all the three curves the total amount of dark matter is the same.}
  \label{fig:APS}
\end{figure}

\begin{figure}[t]
  \begin{center}
    \includegraphics[width=0.7\linewidth]{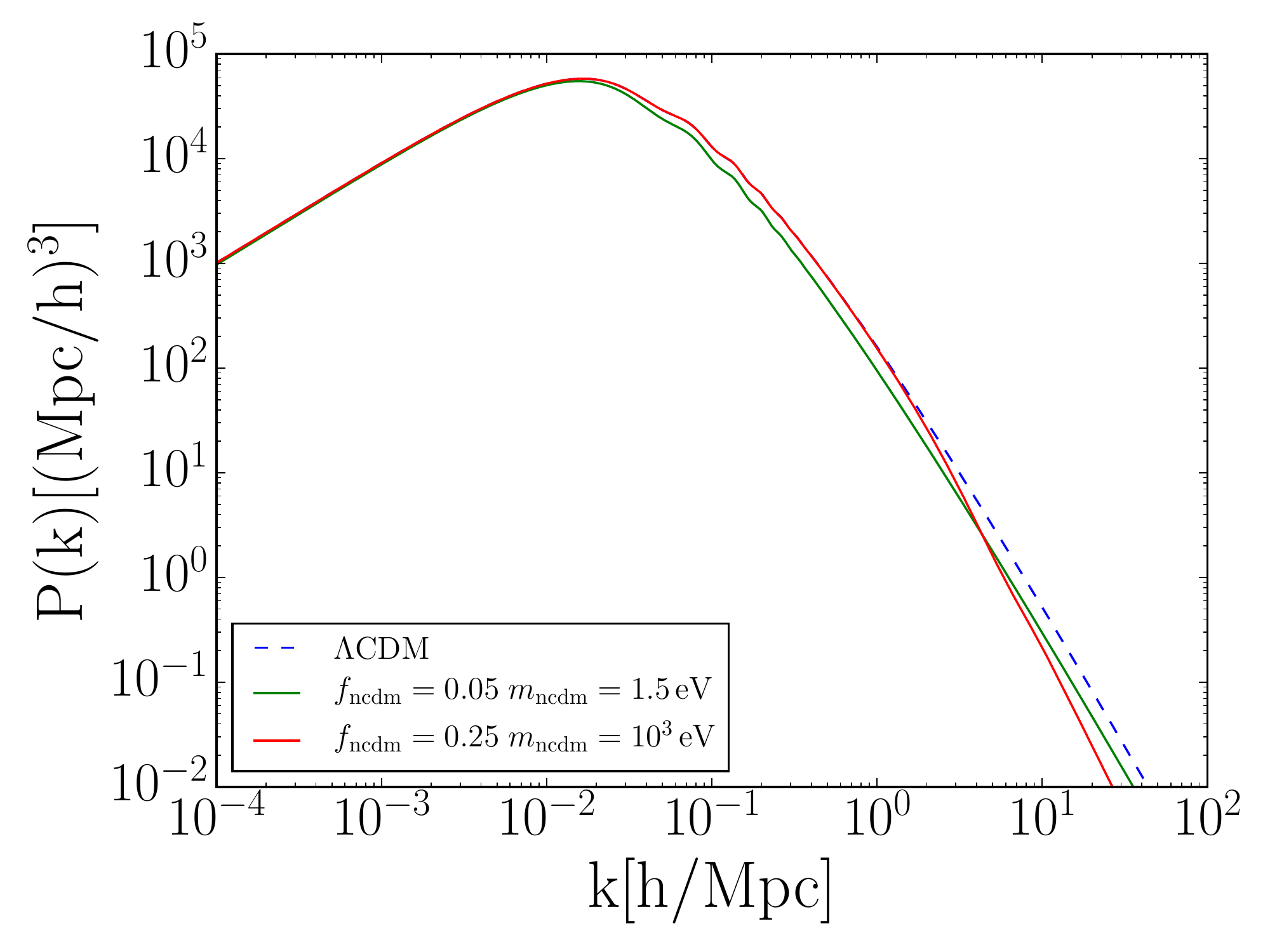}
  \end{center}
  \caption{The same as in Fig.~\ref{fig:APS}, but for the matter power spectrum.}
  \label{fig:MPS}
\end{figure}

Before going to the main results, we first give an impression of how much the standard cosmological picture changes for two exemplary MDM scenarios explored here.  Figures~\ref{fig:APS} and \ref{fig:MPS} show the angular power spectra and the matter power spectra, respectively, obtained by using two models: one with $f_{\rm ncdm} = 0.25$ and $m_{\rm ncdm} = 1$~keV, and a second one with $f_{\rm ncdm} = 0.05$ and $m_{\rm ncdm} = 1.5$~eV.  We show for comparison the predictions for the standard $\Lambda$CDM model, as well as the measurements from the latest data release of Planck satellite~\cite{Ade:2015xua} in the case of the temperature anisotropies. In the figures, all the other cosmological parameters are kept fixed, including the total DM density $\Omega_{\rm dm}$.

Notice that for the case $m_{\rm ncdm} = 1$~keV and $f_{\rm ncdm} = 0.25$ the predictions for the temperature power spectrum are essentially identical to those within the canonical $\Lambda$CDM scenario for most of the scales.  This is due to the fact that CMB physics is basically unaffected by the free streaming nature of a $1$~keV particle that accounts here for $25\%$ of the total DM mass-energy density.  There only exists a tiny difference in the matter power spectrum at very small scales (see Fig.~\ref{fig:MPS}) due to the suppression of the growth of matter perturbations induced by the non-zero velocity dispersions of the NCDM component, even if it is not the dominant one.  Therefore, we expect this point in the parameter space to be allowed by both CMB data and satellite galaxy measurements.  On the other hand, the MDM model with $f_{\rm ncdm} = 0.05$ and $m_{\rm ncdm} = 1.5$~eV gives predictions for the CMB temperature anisotropies that are not compatible with the present CMB data.  This is because the NCDM component behaves as radiation at the decoupling period, enhancing the first acoustic peak height, although the effect will be degenerate with the CDM energy density, in such a way that one could partly compensate this enhancement by increasing the total DM energy density.  However, the suppression in the matter power spectrum for $m_{\rm ncdm} = 1.5$~eV occurs at a much larger scales than before, due to the shorter free-streaming scale, showing a clear discrepancy with power spectrum measurements.

\section{Results}
\label{RES}

\begin{figure}
  \begin{center}
    \includegraphics[width=0.495\linewidth]{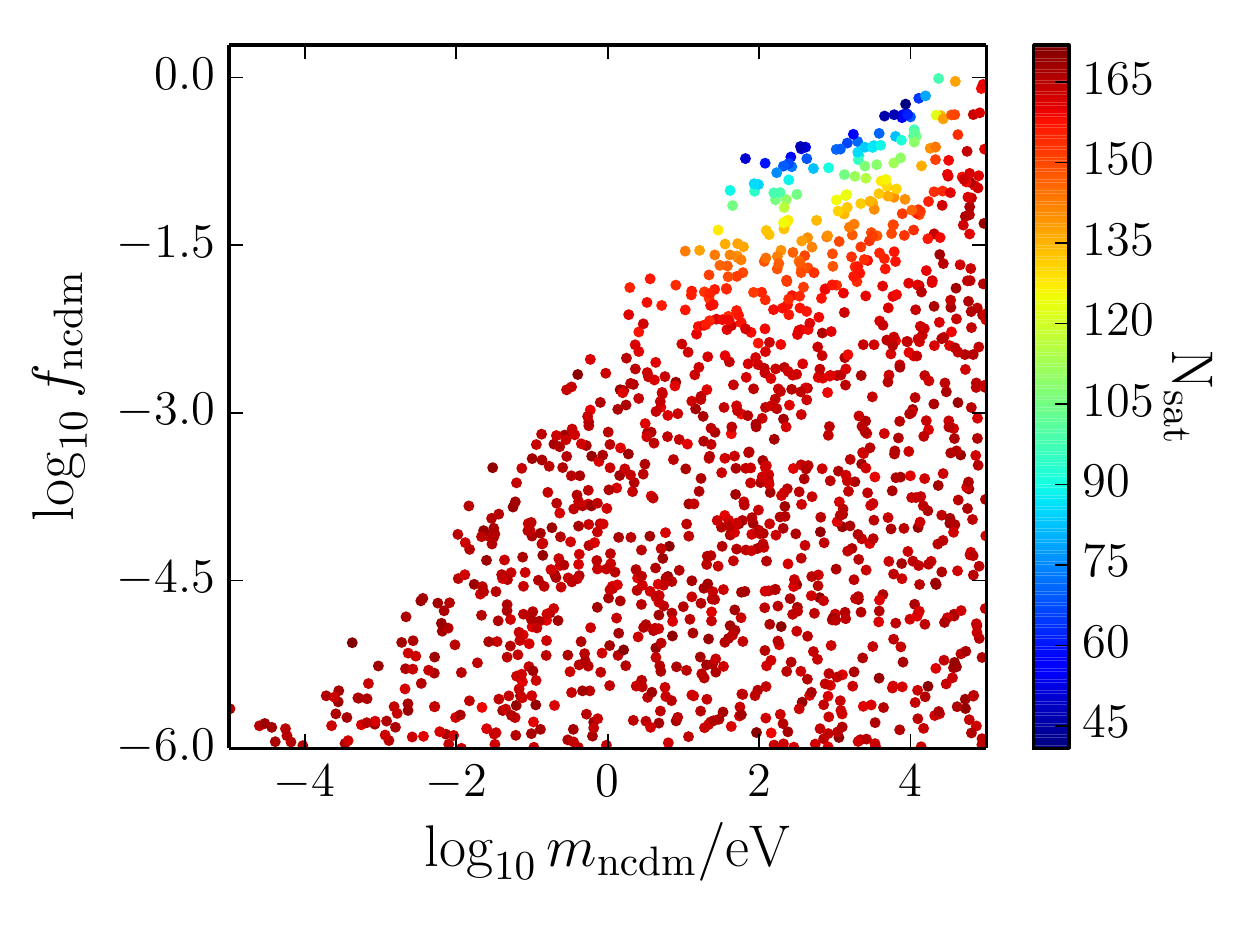}
    \includegraphics[width=0.495\linewidth]{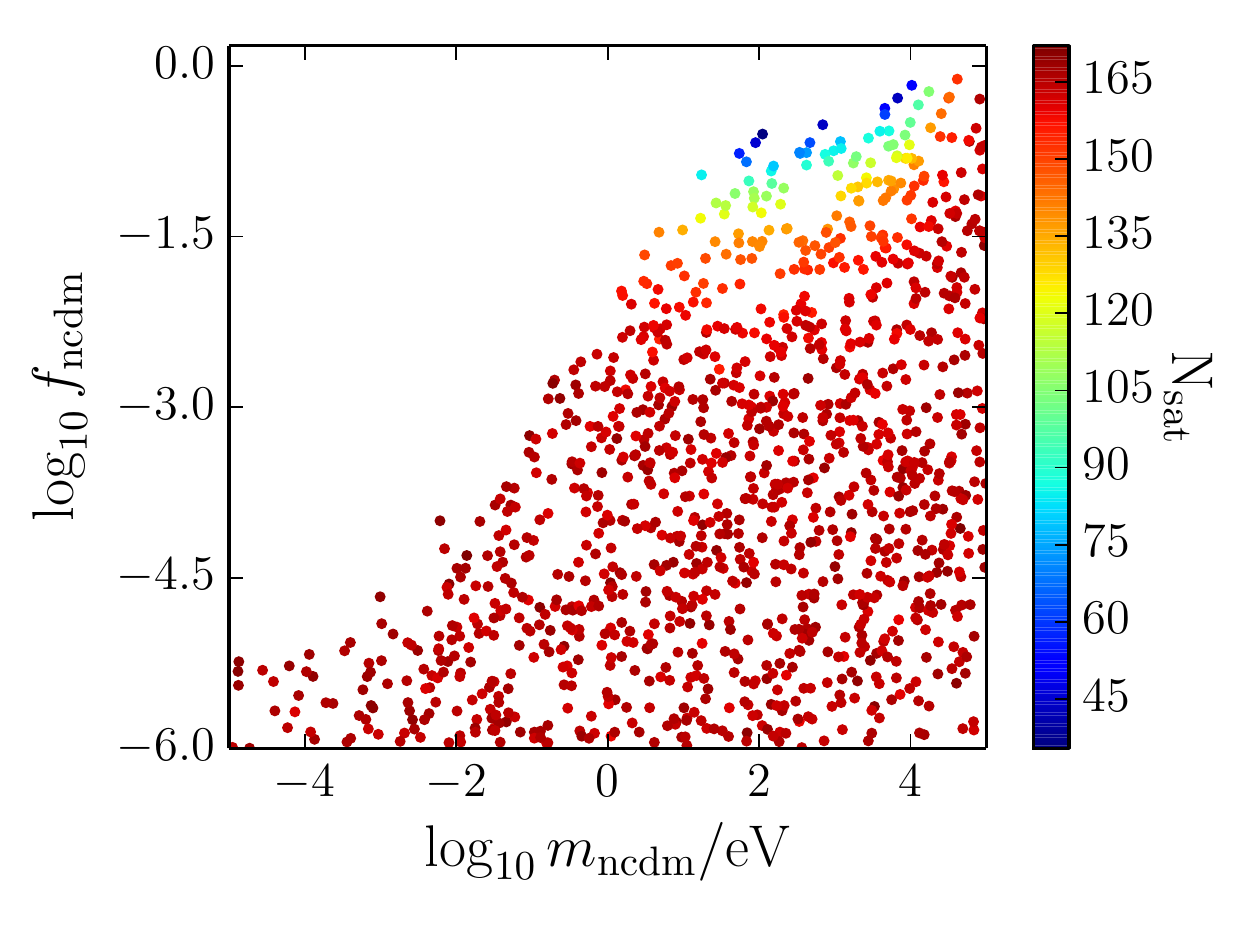}
  \end{center}
  \caption{ Samples in the ($\log_{10}{f_{\rm ncdm}}$, $\log_{10}{m_{\rm ncdm}}$) plane from CMB+SAT+BAO data, colour-coded by the number of the satellite galaxies obtained for the case of a fermionic (left panel) or bosonic (right panel) NCDM candidate.}
  \label{fig:scatterCMB}
\end{figure}

In Fig.~\ref{fig:scatterCMB}, we show, in equal-weight scatter plots, our numerical results as a function of the fraction and mass of the NCDM component. As described above, the results are based on the combination of CMB+SAT+BAO data. Among these data sets, CMB measurements are absolutely required to remove the existing degeneracies among the six $\Lambda$CDM parameters. Furthermore, they also discard at a very high confidence level the region of very low masses for large values of the NCDM fraction $f_{\rm ncdm}$. BAO data also help in further pinning down this NCDM parameter region, albeit in a milder manner than the CMB ones. The satellite likelihood is the one which can break the degeneracy between $f_{\rm ncdm}$ and $m_{\rm ncdm}$, in particular in the observationally  interesting WDM region around the keV scale, and therefore it is crucial for the aim of this work. The samples in Fig.~\ref{fig:scatterCMB} are colour-coded by the predicted number of satellite galaxies.  We find that the number counts of dwarf satellite galaxies start to be noticeably affected for fractions larger than a few percent, and for masses up to around 10 keV.  At 100 keV, the NCDM component behaves essentially as CDM for our purposes.

\medskip

\begin{figure}
  \begin{center}
    \includegraphics[width=0.75\linewidth]{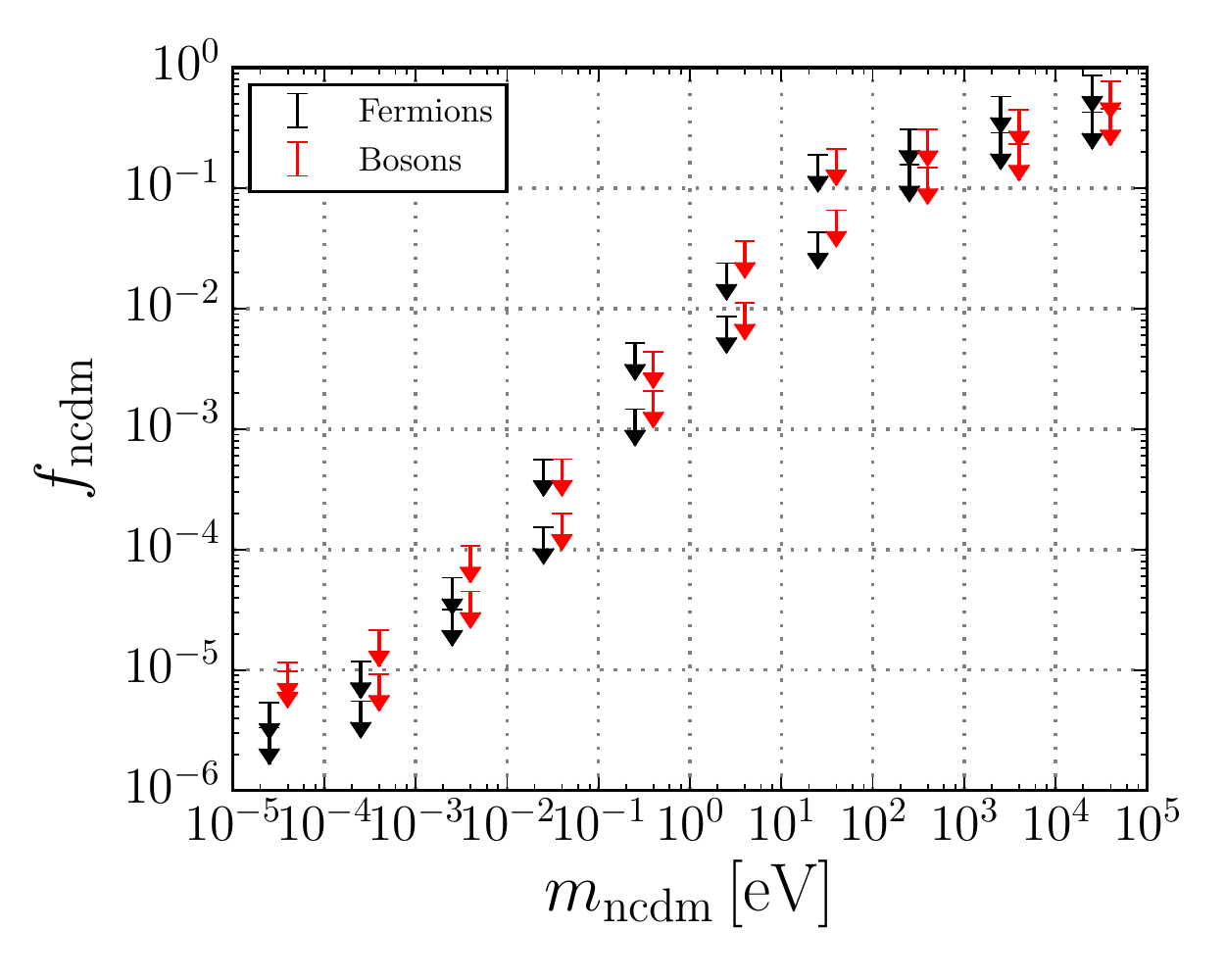}
  \end{center}
  \caption{ $2\sigma$ and $3\sigma$ upper limits on the fraction $f_{\rm ncdm}$ of the NCDM component, obtained for different ranges of masses.  In black (red) we present the results derived from the analysis of the CMB+SAT+BAO datasets for a fermion (boson) NCDM candidate.}
  \label{fig:histogram3}
\end{figure}

In Fig.~\ref{fig:histogram3} we show the main results of this paper.  For each decade in the mass of the NCDM component we show the marginalized upper limits on its fraction $f_{\rm ncdm}$ for both fermions and bosons.  We find that the $95.4\%$CL limits for masses 1--10 keV are $f_{\rm ncdm}\leq0.29$ (0.23) for fermions (bosons), and for masses 10--100 keV they are $f_{\rm ncdm}\leq0.43$ (0.45), respectively.

For large values of the NCDM mass, our limits on its fraction $f_{\rm ncdm}$ are competitive to those existing in the literature; see, \fex, Ref.~\cite{Boyarsky:2008xj}.  In this regime the bounds come mainly from the BAO data and the number of satellites galaxies, since CMB alone is not able to distinguish the heavy NCDM component from a pure CDM one.

On the other hand, in the semi-relativistic regime, 10--100 $\mu$eV, the limits are very strong, $f_{\rm ncdm}\leq3.3\times10^{-6}$ $(9.8\times10^{-6})$ for fermions (bosons).  This is expected, since in the relativistic limit the current cosmological upper bounds on dark radiation apply.  Using Eqs.~\eqref{RHO} and \eqref{NEFF}, indeed, we find for a very light NCDM particle:
\begin{equation}\label{eq:f_darkrad}
  f_{ncdm}
  = \frac{\Omega_{ncdm}}{\Omega_{dm}}
  =  \frac{\Omega_\gamma \, \DNeff}{\Omega_{dm}}
  \approx 10^{-4}\DNeff
  \,.
\end{equation}
Considering that the $95.4\%$CL limits that arise from the standard $\Lambda$CDM+$N_{\rm eff}$ analyses \cite{Ade:2015xua} correspond to $\DNeff\lesssim0.3$, our limits for $f_{\rm ncdm}$ when the mass is around 10--100 $\mu$eV are in reasonable agreement with Eq.~\eqref{eq:f_darkrad}.  The agreement, however, is not complete, especially in the fermion case.  The reason is that the region of parameter space that corresponds to such a case is rather small and the shot noise of the simulation has a significant role, so that even the \texttt{Multinest} algorithm (with the adopted accuracy settings) cannot properly explore it.  The result is that the density of sampled points in the relevant region is not sufficient to obtain exactly the expected constraints corresponding to Eq.~\eqref{eq:f_darkrad}.

\begin{figure}
\begin{center}
    \includegraphics[width=0.495\linewidth]{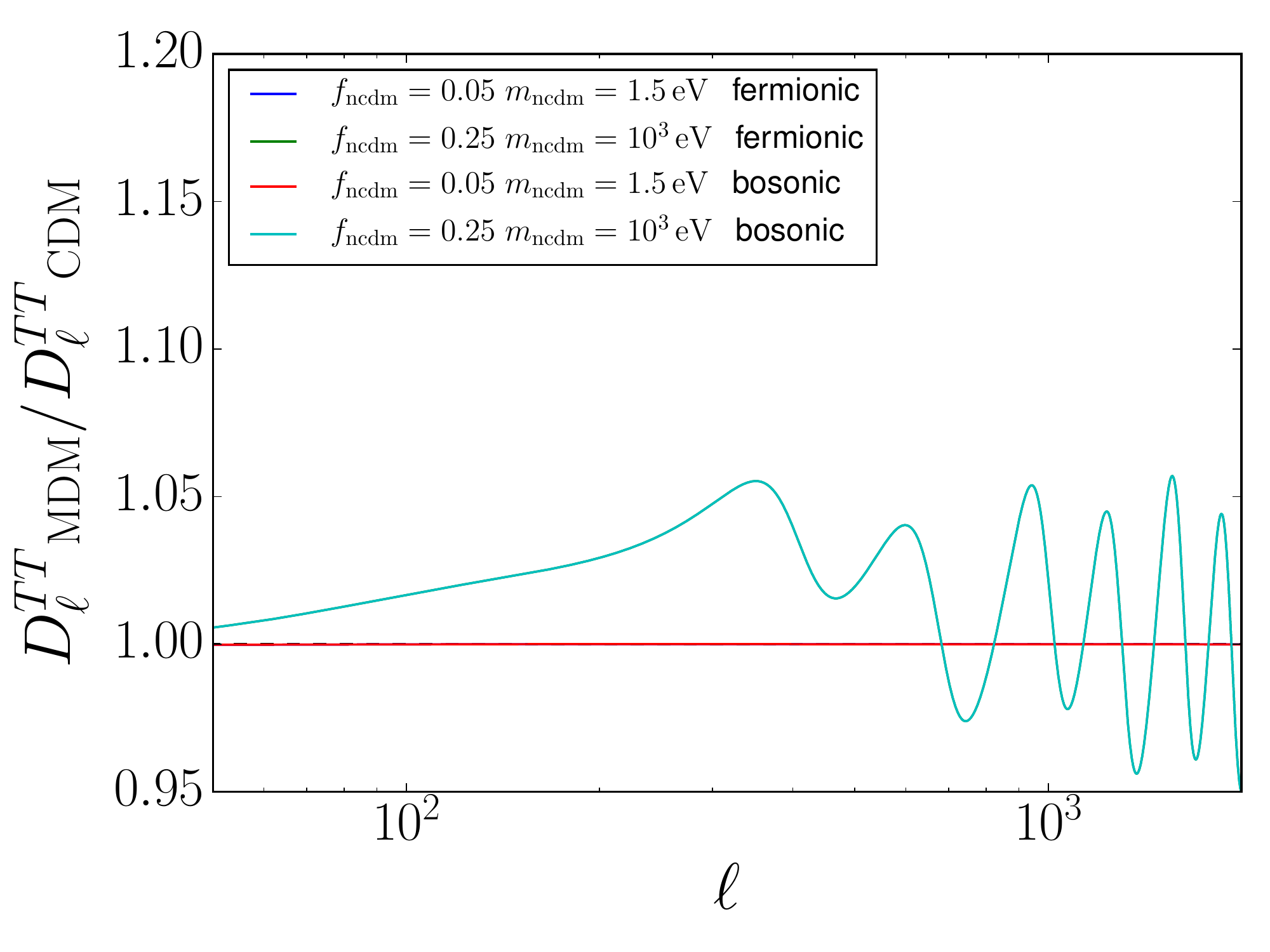}
    \includegraphics[width=0.495\linewidth]{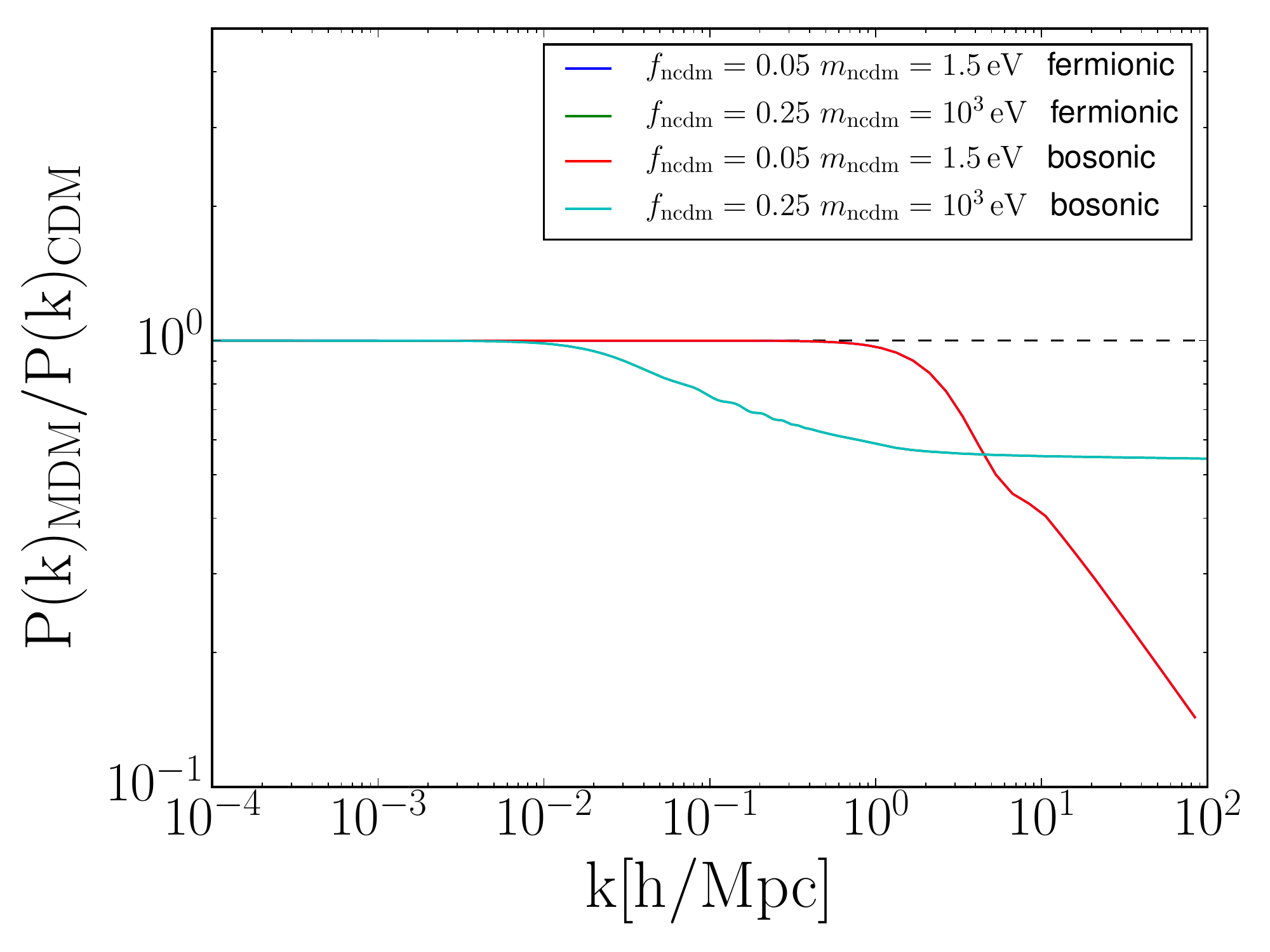}
  \end{center}
  \caption{Left (right) panel: ratios of the temperature anisotropies angular power spectrum (matter power spectrum) for a NCDM component with fraction $f_{\rm ncdm} = 0.05$ ($f_{\rm ncdm} = 0.25$) and mass $m_{\rm ncdm} = 1.5$~eV ($m_{\rm ncdm} = 10^3$~eV) with respect to the $\Lambda$CDM expectations. All the other cosmological parameters are kept fixed. Notice that fermionic and bosonic NCDM predictions are nearly identical.}
  \label{fig:ratios}
\end{figure}

Moreover, as we can see in Fig.~\ref{fig:histogram3}, there are several small unexpected differences between the fermion and the boson case, since, as shown in Fig.~\ref{fig:ratios}, the differences among fermionic and bosonic NCDM candidates are expected to be negligible. The tiny differences in the limits in Fig.~\ref{fig:histogram3} between the fermion and the boson cases are due to the shot noise of the Monte Carlo simulation.

\section{Conclusions}
\label{CON}

In this study, we have explored the cosmological bounds arising from Planck CMB temperature and polarization at low multipoles plus BAO measurements on general cosmological models with a second non-cold dark component (NCDM).  To constrain the NCDM component in the regime where it is warm, we include additional constraints by requiring that the model does not under-predict the number of satellite galaxies observed in the Milky Way.  We adopt phase space distributions for the NCDM that correspond to a component that freezes out while still being relativistic.  We compare results for bosonic and fermionic NCDM.

Our results show that, for the adopted observables, there is not a substantial difference between the allowed regions corresponding to bosonic and fermionic NCDM.  For small NCDM masses, the limit on its fraction relative to the total amount of DM in the universe ($f_{\rm ncdm}$) is around a few times $10^{-5}$.  This limit is approximately close to what one would expect from present constraints on the extra relativistic degrees of freedom $\Delta N_{\rm{eff}}$.  For high values of the NCDM mass, above fractions of keV, the CMB is unable to distinguish among the NCDM and CDM components and one therefore needs to look for independent observables such as the matter power spectrum, that we include through the measurements of the BAO scale.  In that case, we obtain the following $2\sigma$-level upper limits for $f_{\rm ncdm}$: for NCDM particles with mass in the range 1--10 keV, $f_{\rm ncdm}\leq 0.29$ (0.23) in the fermionic (bosonic) case, while for NCDM particles with mass in the range 10--100 keV, $f_{\rm ncdm}\leq 0.43$ (0.45).  For these values of the NCDM mass our limits on its fraction $f_{\rm ncdm}$ are slightly tighter than those existing in the literature (see e.g. Ref.~\cite{Palazzo:2007gz, Boyarsky:2008xj,Poulin:2016nat}).  Forthcoming precise measurements of the matter power spectrum at small scales may be able to further corner mixed DM scenarios.

\acknowledgments

O.M.~is supported by PROMETEO II/2014/050, by the Spanish Grant FPA2014--57816-P of the MINECO, by the MINECO Intramural OEP2010, by the MINECO Grant SEV-2014-0398 and by the European Union's Horizon 2020 research and innovation programme under the Marie Sk{\l o}dowska-Curie grant agreements 690575 and 674896.
The work of S.G.~was supported by the Spanish grants FPA2014-58183-P, Multidark CSD2009-00064 and SEV-2014-0398 (MINECO), and PROMETEOII/2014/084 (Generalitat Valenciana).
The work of R.D., S.A., and C.W.~was supported by NWO through two Vidi grants and partly by University of Amsterdam.


\providecommand{\href}[2]{#2}\begingroup\raggedright\endgroup

\end{document}